\documentclass[prd,aps,reprint,preprintnumbers,showpacs,nofootinbib,superscriptaddress,notitlepage]{revtex4-1} 
\usepackage{slashed}
\usepackage{array}

\usepackage{multirow}
\usepackage{hyperref}
\usepackage{soul}
\usepackage{placeins} 
\usepackage{float}
\newcommand{\Tr}{\operatorname{Tr}}

\usepackage{epsfig}
\usepackage{bbold}

\usepackage{caption}

\usepackage{amssymb,amsthm,amsmath}
\usepackage{xcolor}
\usepackage{textcomp}
\usepackage{verbatim}
\usepackage[normalem]{ulem} 
\usepackage{rotating}   
\begin{document}
	
	\setlength{\parskip}{0pt}

\raggedbottom

\title{Spin-flavor entanglement in $\Lambda_b \to \Lambda D$ and weak phase extraction}

\author{Yong Du}\email{yongdu5@impcas.ac.cn} 
\affiliation{Institute of Modern Physics, Chinese Academy of Sciences, Lanzhou 730000, China} 
\affiliation{School of Nuclear Science and Technology, University of Chinese Academy of Sciences,\\
	19A Yuquan Road, Beijing 100049, China}

\author{Chao-Qiang Geng}\email{cqgeng@ucas.ac.cn} 
\affiliation{School of Fundamental Physics and Mathematical Sciences, Hangzhou Institute for Advanced Study, UCAS, Hangzhou 310024, China}

\author{Xiao-Gang He}\email{hexg@sjtu.edu.cn}

\affiliation{State Key Laboratory of Dark Matter Physics, Tsung-Dao Lee Institute and School of Physics and Astronomy, Shanghai Jiao Tong University, Shanghai 201210, China}
\affiliation{Key Laboratory for Particle Astrophysics and Cosmology (MOE) \& Shanghai Key Laboratory for Particle Physics and Cosmology, Tsung-Dao Lee Institute and School of Physics and Astronomy, Shanghai Jiao Tong University, Shanghai 201210, China}

\author{Chia-Wei Liu}\email{chiaweiliu@ucas.ac.cn}
\affiliation{School of Fundamental Physics and Mathematical Sciences, Hangzhou Institute for Advanced Study, UCAS, Hangzhou 310024, China}

\author{Sheng-Lin Liu}\email{liushenglin22@mails.ucas.ac.cn}

\affiliation{School of Fundamental Physics and Mathematical Sciences, Hangzhou Institute for Advanced Study, UCAS, Hangzhou 310024, China}
\affiliation{Institute of Theoretical Physics, UCAS, Beijing 100190, China}
\affiliation{University of Chinese Academy of Sciences, Beijing 100190, China}

\author{Xin-Yi Liu}\email{liuxinyi24@mails.ucas.ac.cn}

\affiliation{School of Fundamental Physics and Mathematical Sciences, Hangzhou Institute for Advanced Study, UCAS, Hangzhou 310024, China}
\affiliation{Institute of Theoretical Physics, UCAS, Beijing 100190, China}
\affiliation{University of Chinese Academy of Sciences, Beijing 100190, China}

\begin{abstract}

We identify a new spin--flavor entanglement structure in $\Lambda_b\to\Lambda D$ decays, formed by the correlation between the $\Lambda$ spin and the  $D$ flavor states ($D=D^0,\overline D^0,D_1,D_2$). The entanglement information is encoded in the decay rates and Lee--Yang parameters of the four neutral-$D$ modes.  We then show that the same spin--flavor structure provides a new method to determine the weak phase $\gamma$, a key angle of the Cabibbo--Kobayashi--Maskawa unitarity triangle. We find that the experimental uncertainty scales as $\sigma_\gamma\propto 1/{\cal C}$, where ${\cal C}$ is the Wootters concurrence, thereby quantitatively relating the precision of the weak-phase extraction to the amount of spin--flavor entanglement. 

\end{abstract}

\date{\today}
 \maketitle
\section{Introduction}
Entanglement provides a useful organizing principle for collider observables,
from $t\bar t$ quantum
tomography and entanglement at the LHC~\cite{Afik:2020onf,ATLAS:2023fsd,
CMS:2024pts,Gu:2025rg,Han:2024kct} to fermion-pair production at lepton
colliders~\cite{Fang:2026ddi,Cheng:2025xux,Cao:2025qua}.  Related spin-correlation systems include
$\Lambda\bar\Lambda$ production at BESIII~\cite{BESIII:2018cnd,Perotti:2018wxm,
	Wu:2024bne}
and recent baryon-pair entanglement, decay-distillation, and confinement
studies~\cite{Du:2024sly,Lin:2025eci,Chen:2026oaf,Feng:2025ryr,STAR:2025njp,
	Oliva:2026qsc},
as well as neutral-$B$ entanglement in $\Upsilon(4S)$
decays~\cite{Belle:2007EPR}.  In this work, we identify a different form of entanglement,
the spin--flavor 
entanglement, in $\Lambda_b\to\Lambda D$, where
$D=D^0,\overline{D}^0,D_1,D_2$.  The neutral-$D$ flavor and the $\Lambda$ spin form a two-state spin--flavor
system whose density matrix can be probed through the decay rates and
Lee--Yang parameters~\cite{LeeYang:1957}.

\begin{figure}[H]
	\centering
	\includegraphics[width=\columnwidth]{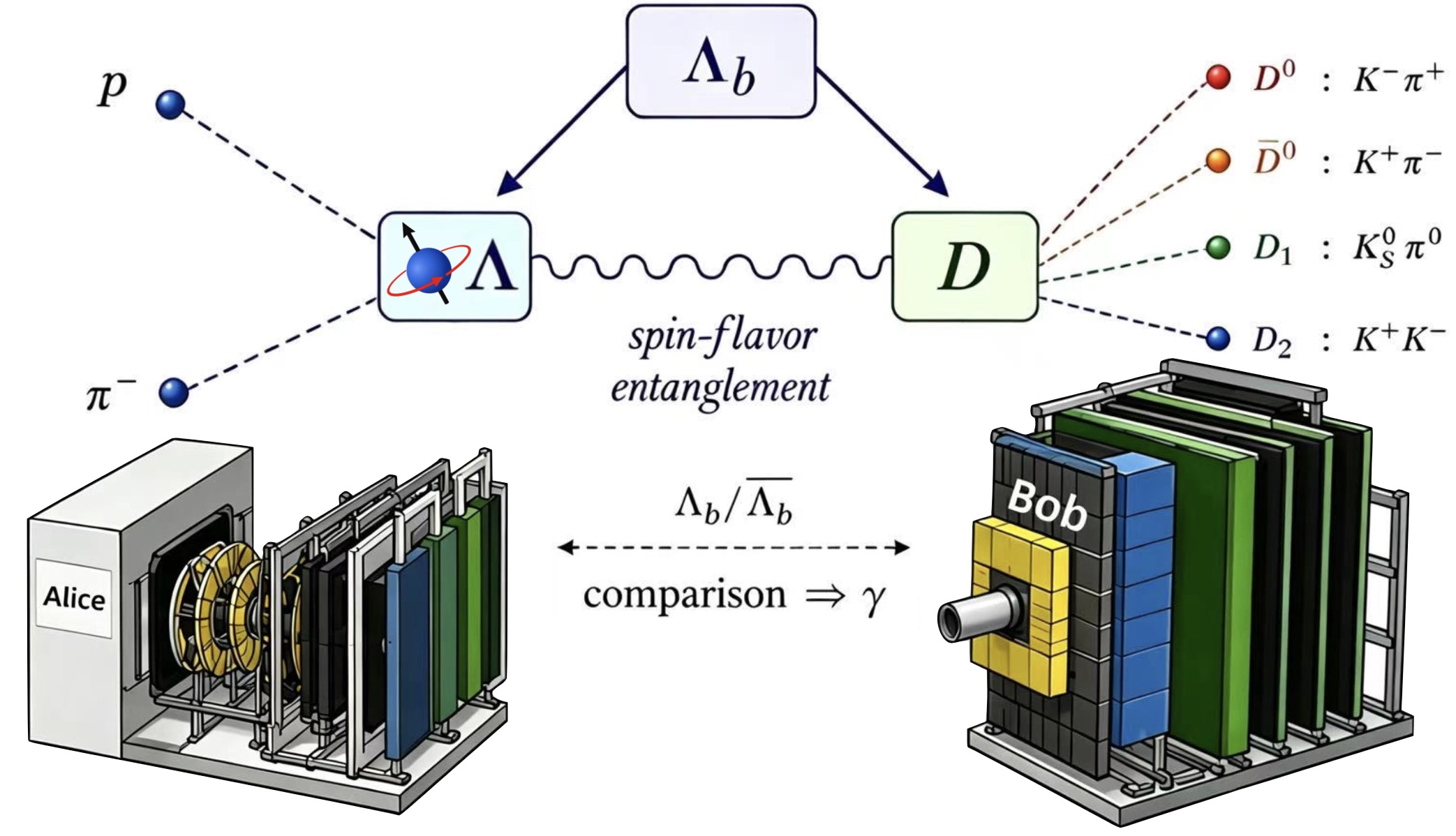}
	\caption{ Schematic plot for spin--flavor entanglement in $\Lambda_b \to \Lambda D$, where Alice measures the spin and Bob the flavor.  
	}
	\label{fig:abstract}
\end{figure}

The weak phase \(\gamma\) is the Cabibbo--Kobayashi--Maskawa (CKM)
angle~\cite{Cabibbo:1963yz,Kobayashi:1973fv}, defined as the argument of the CKM matrix element $V_{ij}$ ratio:
\begin{equation}
	\gamma
	\equiv
	\arg\left[
	-\frac{V_{ud}V_{ub}^*}{V_{cd}V_{cb}^*}
	\right],
\end{equation}
which appears in the charged-current flavor structure of the Standard
Model.  Since \(\gamma\) can be determined from tree-level amplitudes, its
theoretical uncertainty is extremely small compared with current and planned
experimental precision~\cite{PDG:2024,Brod:2013sga}, as demonstrated by the
methods employed in \(B\to D\pi\) and \(B\to DK\)
decays~\cite{Gronau:1991dp,Gronau:1991dk,Atwood:1996ci,Atwood:2000ck,
	Giri:2003ty,LHCb:2021sqa,BelleBelleII:2024}.  The same interference mechanism
is present in \(\Lambda_b\to\Lambda D\): the \(V_{ub}V_{cs}^*\) contribution
is embedded in \(\Lambda_b\to\Lambda D^0\), while the \(V_{cb}V_{us}^*\)
contribution is embedded in \(\Lambda_b\to\Lambda\overline{D}^0\).  Thus, the
two flavor-tagged modes carry different CKM phases and interfere when the
neutral \(D\) is reconstructed in the CP eigenstates
\(|D_1\rangle=(|D^0\rangle+|\overline{D}^0\rangle)/\sqrt{2}\) and
\(|D_2\rangle=(|D^0\rangle-|\overline{D}^0\rangle)/\sqrt{2}\).  The weak-phase information is encoded in the entanglement between the
$\Lambda$ spin and the neutral-$D$ flavor.
	We study this entanglement, depicted in Figure~\ref{fig:abstract}, and its role
	in weak-phase extraction.

	\section{
		SPIN--FLAVOR DENSITY MATRIX AND WEAK-PHASE SENSITIVITY 
	}
	 
At the quark level, \(\Lambda_b\to\Lambda D\) receives two tree-level
charged-current contributions.  The effective Hamiltonian is written as~\cite{Buchalla:1995vs, Geng:2022osc}  
\begin{equation}\label{eq:heff_intro}
\begin{aligned}
{\cal H}_{\rm eff}
= {}&
\frac{G_F}{\sqrt2}V_{cb}V_{us}^* 
\bigg[
\left(
C_1 Q_1^{(c)}+C_2 Q_2^{(c)}
\right)
\\
+&
\left|\frac{ V_{ub}V_{cs}^*}{V_{cb}V_{us}^* } \right|e^{-i\gamma}
\left(
C_1 Q_1^{(u)}+C_2 Q_2^{(u)}
\right)
\bigg]
+{\rm h.c.}\, .
\end{aligned}
\end{equation}
Here \(G_F\) is the Fermi constant, \(C_i\) are Wilson coefficients, and
\(Q_i^{(c)}\) and \(Q_i^{(u)}\) denote the four-quark operators for the
\(b\to c\bar u s\) and \(b\to u\bar c s\) transitions, respectively.  Since
the initial baryon \(\Lambda_b\) has spin \(1/2\), the corresponding hadronic
matrix elements are most conveniently described in the \(J=1/2\) partial-wave
basis with fixed \(J_z\); the \(\Lambda\) and \(J=1/2\) labels are suppressed
below.  Details of the quark-operator definitions, the eigenstate
normalization ${\cal N}$, and the relation to the usual Dirac spinor expansion are given
in Appendix~\ref{appa}.  The effective Hamiltonian in
Eq.~\eqref{eq:heff_intro} induces the spin--flavor state
\begin{equation}\label{heli_exp}
	|\psi, {J_z}  \rangle
	=
	\frac{1}{{\cal N}}
	\left[
	e^{-i\gamma}
	|\overline{D}^0\rangle\otimes \chi_{\overline{D}}
	+
	|D^0\rangle\otimes \chi_D
	\right],
\end{equation}
where
\begin{equation}
	\chi_{\overline{D}}
	=
	\begin{pmatrix}
		S_{\overline{D}^0}+P_{\overline{D}^0}
		\\
		S_{\overline{D}^0}-P_{\overline{D}^0}
	\end{pmatrix},
	\qquad
	\chi_D
	=
	\begin{pmatrix}
		S_{D^0}+P_{D^0}
		\\
		S_{D^0}-P_{D^0}
	\end{pmatrix},
\end{equation}
 and we use \(S_D\) and \(P_D\) to denote the \(S\)- and \(P\)-wave amplitudes for
\(\Lambda_b\to\Lambda D\), respectively.  The two-component spinors
\(\chi_{\overline D}\) and \(\chi_D\) are written in the
\(\Lambda\)-helicity angular-momentum basis,
with \(|J_z,\lambda=+\tfrac12;\Lambda D\rangle=(1,0)^T\) and
\(|J_z,\lambda=-\tfrac12;\Lambda D\rangle=(0,1)^T\).

Noting that $|\psi, J_z\rangle$ lives in the tensor product of the
neutral-\(D\) flavor and \(\Lambda\) spin spaces, we define the Lee--Yang
operators

\begin{equation}
\begin{aligned}
	\hat{\alpha}
	&= 2 \hat{k}\cdot \vec{s},
\quad 	\hat{\beta}
=
2
\left(
\vec{J}\times \hat{k}
\right)\cdot \vec{s},
	\\
	\hat{\gamma}
	&=
	2\left(
	\vec{J}
	-
	\hat{k}\,\hat{k}\cdot \vec{J}
	\right)\cdot \vec{s}, 
\end{aligned}
\end{equation}
where $\hat{k}$ is the unit vector along the $\Lambda$ momentum in the
$\Lambda_b$ rest frame, and $\vec{s}$ is the spin operator of the final-state
$\Lambda$~\cite{spin}. 
These Lee--Yang operators act only on the helicity spinors
\(\chi_D\) and \(\chi_{\overline{D}}\).  In the $\Lambda$-helicity basis, their action is
equivalent to
\(\hat\alpha\to\sigma_z\), \(\hat\gamma\to\sigma_x\), and
\(\hat\beta\to-\sigma_y\) as demonstrated in Appendix~\ref{app:lee_yang_binary}.

Furthermore, we identify \(\overline{D}^0\) and
\(D^0\) as \((1,0)^T\) and \((0,1)^T\) in flavor space, such that
\(\hat O_F\) and \(\hat O_{12}\) play the roles of
\(\boldsymbol\sigma_z\) and \(\boldsymbol\sigma_x\) in flavor space:
\begin{equation}\label{flavor}
\begin{aligned}
	\hat{O}_F
	&=
	| \overline{D}^0 \rangle \langle \overline{D}^0 |
	-
	| D^0 \rangle \langle D^0 |,
	\\
	\hat{O}_{12}
	&=
	| D_1 \rangle \langle D_1 |
	-
	| D_2 \rangle \langle D_2 | .
\end{aligned}
\end{equation}
The flavor direction \(\boldsymbol\sigma_y\) is not directly observable
experimentally, since the states \(D^0\pm i\overline{D}^0\) cannot be tagged.
In theory, it can nevertheless be obtained from
\(\boldsymbol\sigma_y=i\boldsymbol\sigma_x\boldsymbol\sigma_z\).

The full spin--flavor density matrix is
\(\rho_{J_z}=|\psi, {J_z}\rangle\langle\psi, {J_z}|\).
The helicity density matrix for a given neutral-\(D\) component is obtained
by projecting onto that flavor state:
\begin{equation}
	\rho_{J_z,D}
	=
	\Tr_{ D}
	\left(
	\rho_{J_z}\,\hat O_D
	\right),
\end{equation}
and \(\Tr_{  D}\) denotes the partial trace over the two-dimensional
neutral-\(D\) flavor space.    The relevant
projectors are
\(\hat O_{\overline{D}^0,D^0}=(1\pm\hat O_F)/2\) and
\(\hat O_{D_1,D_2}=(1\pm\hat O_{12})/2\), where the upper and lower signs
refer to the first and second labels, respectively.

A further trace over the
helicity space gives the Lee--Yang parameters
\begin{equation}
	\xi_D
	=
	\frac{
		\Tr_{\lambda}(\rho_{J_z,D}\hat\xi)
	}{
		\Tr_{\lambda}(\rho_{J_z,D})
	},
	\qquad
	 \xi\in\{ \alpha, \beta, \gamma\}.
\end{equation}
Thus the hatted quantities \(\hat\xi\), or equivalently the channel-projected
operators \(\hat{\xi}_D =\hat O_D\hat\xi\), are Lee--Yang operators,
while their unhatted expectation values
\(\ \langle\hat{\xi}_D\rangle \in\{\alpha_D,\beta_D,\gamma_D\}\)
are Lee--Yang parameters: 

\begin{equation}
\begin{aligned}
	\alpha_D
	&=
	\frac{2\,\mathrm{Re}(S_D^*P_D)}
	{|S_D|^2+|P_D|^2},
	\quad 
	\beta_D
	 =
	\frac{2\,\mathrm{Im}(S_D^*P_D)}
	{|S_D|^2+|P_D|^2},
	\\
	\gamma_D
	&=
	\frac{|S_D|^2-|P_D|^2}
	{|S_D|^2+|P_D|^2}.
\end{aligned}
\end{equation}
Note that $\gamma_D$ here is a Lee--Yang parameter and should not be confused
with the weak phase. 

The degree of spin--flavor entanglement is measured by the Wootters
concurrence~\cite{Wootters:1997id}.
For the pure state in Eq.~\eqref{heli_exp}, it is given by
${\cal C}=2|\chi_D^T i\sigma_y \chi_{\overline D}|/{\cal N}^2$, or equivalently
\begin{equation} \label{crit}
\begin{aligned}
	{\cal C}
	&=
	\sqrt{
		\frac{1-R_F^2}{2}
		\left(
			1-\sum_{\xi=\alpha,\beta,\gamma}
			\xi_{\overline D^0}\xi_{D^0}
		\right)
	} \, .
\end{aligned}
\end{equation} 
Here \(R_F=\langle \hat O_F\rangle\) and
\(R_{12}=\langle \hat O_{12}\rangle\) denote the asymmetries in the flavor and
CP-eigenstate bases, respectively,
\begin{equation}
	R_F
	=
	\frac{
		\Gamma_{\overline{D}^0}
		-
		\Gamma_{D^0}
	}{
		\Gamma_{\overline{D}^0}
		+
		\Gamma_{D^0}
	},
	\qquad
	R_{12}
	=
	\frac{
		\Gamma_{D_1}
		-
		\Gamma_{D_2}
	}{
		\Gamma_{D_1}
		+
		\Gamma_{D_2}
	},
\end{equation}
where \(\Gamma_D=\Gamma(\Lambda_b\to\Lambda D)\). 
For \(R_{F,12}\neq 1\), Eq.~\eqref{crit} indicates that the concurrence
vanishes only if the two Lee--Yang vectors satisfy
$\xi_{\overline{D}^0 } = \xi _{D^0}$.   
Because there is only one weak
phase in the \(D^0\) and \(\overline{D}^0\) amplitudes, there are no direct
CP-odd observables in the flavor-tagged modes
\(\Lambda_b\to\Lambda D^0\) and
\(\Lambda_b\to\Lambda\overline{D}^0\).  Hence
\(
R_F=\overline R_F,\,
\xi_{D^0}\xi_{\overline{D}^0} =\overline\xi_{\overline{D}^0}\overline\xi_{D^0}
\),
and from Eq.~\eqref{crit}, the CP-conjugate process has the same
concurrence.

The observables \(R_{12}\) and \(\overline R_{12}\) provide two constraints, but
they involve three unknowns:
\(\left|\chi_{\overline{D}}^{\dagger}\chi_D\right|\),
\(\arg(\chi_{\overline{D}}^{\dagger}\chi_D)\), and \(\gamma\). Hence the weak phase
cannot be determined from the rate asymmetries alone. The situation remains
the same even if \(\alpha_D\) is determined for all \(D\) modes,  although one
could still obtain a range for \({\cal C}\) in this case  as shown in Eq.~\eqref{interval}.  Thus, at least
two independent Lee--Yang parameters must be measured in order to
overconstrain the system and extract the weak phase.

As a concrete illustration, one may use the projective coordinate of the
reconstructed $\Lambda$-helicity spinor 
\begin{equation}
	z_D
	\equiv
	\frac{\alpha_D+i\beta_D}{1+\gamma_D}.
\end{equation}
The two CP-eigenstate channels then give
\begin{equation}\label{eq15}
	\gamma 
	=
	\frac12
	\arg
	\left[
	\frac{
		\left(z_{\overline{D}^0}-z_{D_{1,2}}\right)
		\left(\bar z_{D^0}-\bar z_{D_{1,2}}\right)
	}{
		\left(z_{D^0}-z_{D_{1,2}}\right)
		\left(\bar z_{\overline{D}^0}-\bar z_{D_{1,2}}\right)
	}
	\right],
\end{equation}
up to $\pi$, with $\bar z_D$ defined from $\bar \xi_D$ in the same way.
The quantities inside the argument are cross ratios and are therefore
invariant under a common $SU(2)$ rotation of the helicity basis. The same
basis change acts on the Lee--Yang vector through the double-cover map,
$ 
    	(\gamma_D,-\beta_D,\alpha_D)^T
	\to
	R (\gamma_D,-\beta_D,\alpha_D)^T,
$ 
with $R \in SO(3)$. 
This basis freedom indicates that $\gamma$ is overconstrained and  can thus be  extracted without knowing the branching
fractions. To use the full information and obtain the strongest constraint,
the most direct approach is to determine $\gamma$ together with the four
complex hadronic amplitudes $S_{\overline{D}^0}$, $P_{\overline{D}^0}$,
$S_{D^0}$, and $P_{D^0}$ in a simultaneous fit to the measured rates and
Lee--Yang parameters.  



To estimate the uncertainty in $\gamma$ extraction, we separate the
contributions into off-diagonal and diagonal parts, associated respectively
with the $SU(2)$-invariant $\chi_{\overline D}^T i\sigma_y \chi_D$ and
$\chi_D^\dagger\chi_{\overline D}$.  
The off-diagonal part determines $\gamma$ with an uncertainty
$\sigma_{\gamma,{\rm off}}\simeq \sigma_\xi/{\cal C}$, where $\sigma_\xi$
is an effective Lee--Yang uncertainty for the rate-weighted transverse
coherence.  
The remaining
diagonal information gives
only branch-dependent constraints. 
We summarize its contribution by a
dimensionless, non-universal function $f_{\rm br}(R_F,{\cal C})$, together
with an effective Lee--Yang uncertainty $\sigma'_\xi$ for the diagonal branch
information. These relations are derived in
Appendix~\ref{app:gamma_density_matrix}. Combining the two contributions
through an inverse-variance estimate gives
\begin{equation}\label{main}
	\sigma_\gamma^{\rm full}
	\simeq
	\frac{1}{{\cal C}}
	\left[
	\frac{1}{\sigma_\xi^2}
	+
	\frac{1}{
		f_{\rm br}^2(R_F,{\cal C})(\sigma'_\xi)^2
	}
	\right]^{-1/2}.
\end{equation}
The numerical coefficient in the square bracket is analysis dependent and can
be modified by rate uncertainties, correlations, and the branch-selection
prescription, but the leading behavior
$\sigma_\gamma^{\rm full}\propto1/{\cal C}$ as ${\cal C}\to0$ is unchanged.
Note that a similar ill-defined behavior occurs in Eq.~\eqref{eq15}, since
$z_{D^0}=z_{D_{1,2}}$ and $\bar z_{D^0}=\bar z_{D_{1,2}}$ in the limit
${\cal C}=0$.

We estimate the statistical resolution of the weak phase with an
observable-level Monte Carlo simulation. For each input concurrence
${\cal C}$, we generate random $S$- and $P$-wave amplitudes at fixed
${\cal C}$, from which the partial widths and Lee--Yang parameters of the four
neutral-$D$ modes are obtained. For the full Run-3 data set of LHCb, we
take~\cite{LHCb:2016qpe,Vecchi:2018bjg,LHCb:2019fns,Rui:2026ihu}
\begin{equation}\label{number}
	N_{\rm tot}
	=
	N_{\Lambda_b}\,
	{\rm Br}(\Lambda_b^0\to\Lambda D)
	\simeq
	3.6 \times10^7,
\end{equation}
as the total pre-reconstruction yield. The reconstruction efficiencies vary
with the $D$-flavor channel and are taken from Ref.~\cite{Geng:2022osc} and
references therein. The channel Lee--Yang operators $\hat{\xi}_D$ are
dichotomic spin observables with outcomes $\pm1$, as shown in
Appendix~\ref{app:lee_yang_binary}. Their statistical uncertainties therefore
follow from finite samples, as given in Eq.~\eqref{uncertainties}.  For illustration, we take
$p_z=5\%$ as our benchmark. The rate asymmetries $R_F$ and $R_{12}$ are obtained
from the efficiency-corrected yields, with uncertainties determined by the
observed event numbers. We apply the same procedure to the CP-conjugate modes
and  reconstruct $\gamma$ in the range
$0\leq\gamma\leq\pi$.

\begin{figure} 
	\centering
	\includegraphics[width=0.8 \columnwidth]{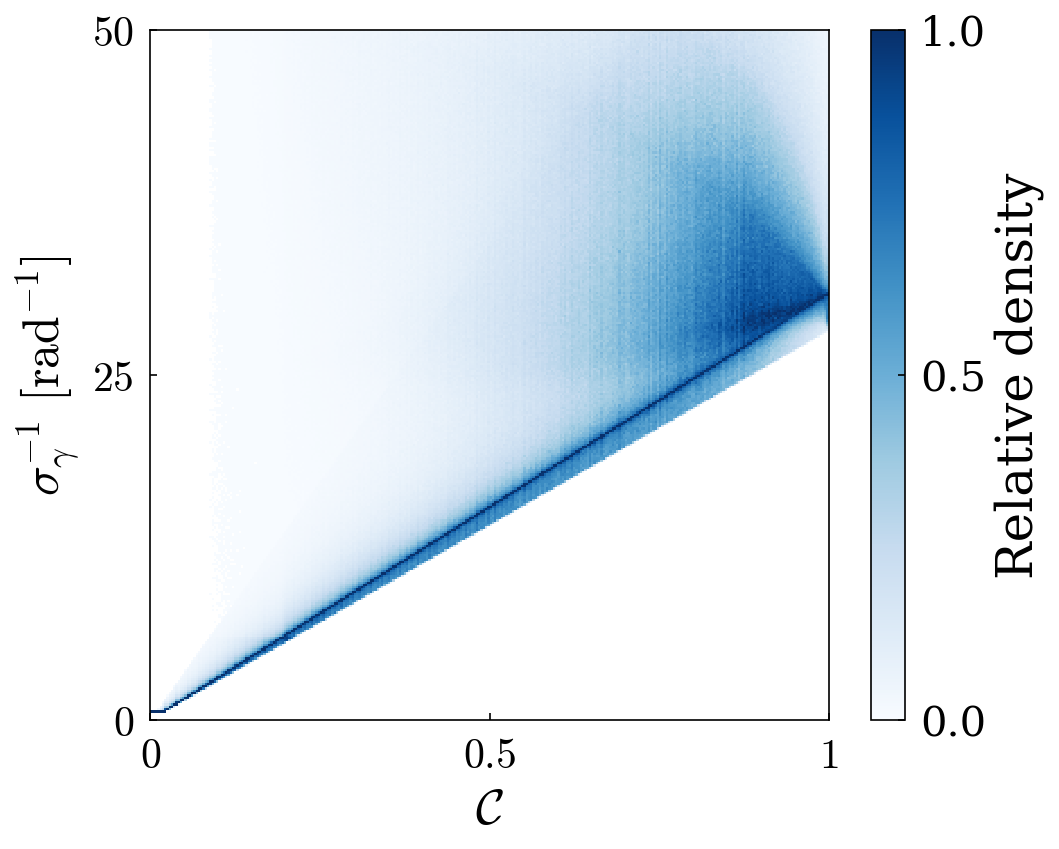}
	\caption{
		Relative-density distribution of $(\sigma_\gamma)^{-1}$ versus the spin--flavor concurrence ${\cal C}$, with $(\sigma_\gamma)^{-1}$ given in ${\rm rad}^{-1}$. The lower $y$ cutoff reflects $\sigma_\gamma \leq \pi/2 $. 
	}
	\label{fig:gamma_resolution_density}
\end{figure}

Figure~\ref{fig:gamma_resolution_density} shows the relative-density
distribution of the scaled inverse resolution $(\sigma_\gamma)^{-1}$ as a
function of ${\cal C}$. At ${\cal C}=0$, the two flavor-tagged helicity
spinors become parallel, and the reconstruction of $\gamma$ becomes
ill-conditioned, consistent with the scaling
$\sigma_\gamma\propto1/{\cal C}$ in Eq.~\eqref{main}. The effective lower
bound on the resolution is controlled by $\sigma_\xi/{\cal C}$. When the
off-diagonal sensitivity is weaker than the branch-cut uncertainty from the
diagonal spinor information, the diagonal contribution becomes ineffective and
the full uncertainty follows $\sigma_\gamma^{\rm full}\simeq\sigma_\xi/{\cal C}$.
Overall, the upward shift of the density band with increasing ${\cal C}$ shows
that spin--flavor entanglement is the dominant driver of weak-phase
sensitivity.

Using perturbative quantum chromodynamics (pQCD) input~\cite{Rui:2026ihu}, we find ${\cal C}=0.18$, corresponding to $\sigma_\gamma\simeq0.6^\circ$ for $p_z=1$ and $\sigma_\gamma\simeq11^\circ$ for $p_z=5\%$ as inferred from Figure\,\ref{fig:gamma_resolution_density} based on our simulation. These values should be compared
with the indirect CKM-unitarity-fit result,
$\gamma=(66.3^{+0.7}_{-1.9})^\circ$, obtained by
CKMfitter~\cite{Hocker:2001xe}. Thus, the baryonic mode is best regarded as a
complementary probe of $\gamma$, providing an independent spin--flavor test of
the weak phase rather than a direct competitor to existing precision methods.


	\section{LOCAL MOMENTUM DENSITY MATRIX AND OBSERVABLE CONCURRENCE
    }
	Motivated by the role of concurrence in extracting the weak phase, we
	investigate the spin--flavor entanglement in $\Lambda_b\to\Lambda D$ in
	this section. While the discussion below focuses on
	$\Lambda_b\to\Lambda D$, the same setup applies to its CP counterpart. We
	start by projecting out the three-momentum eigenstates from the
	spin--flavor state discussed above, following the standard \(S\)- and
	\(P\)-wave parametrization of nonleptonic baryon decays~\cite{LeeYang:1957}: 
	\begin{equation}\label{M_kin}
		\boldsymbol{\cal M}
		=
		i\,\overline u
		\left(
		\boldsymbol S
		-
		\frac{\boldsymbol P}{\kappa}\gamma_5
		\right)
		u_b ,
	\end{equation}
	with \(u_{(b)}\) the Dirac spinors of \(\Lambda_{(b)}\),
	\(\boldsymbol S=(S_{\overline{D}^0},S_{D^0})^T\),
	\(\boldsymbol P=(P_{\overline{D}^0},P_{D^0})^T\), and
	\(\kappa=\sqrt{(E-m_\Lambda)/(E+m_\Lambda)}\). Here \(E\) is the energy of
	\(\Lambda\) in the \(\Lambda_b\) rest frame, and the bold symbols
	\(\boldsymbol{\cal M}\), \(\boldsymbol S\), and \(\boldsymbol P\) denote
	objects in the \(D\)-flavor space.
	The generalized \(2\times2\) differential partial-width matrix
	\(\boldsymbol\Gamma\) in this \(D\)-flavor space is then written as
	\begin{widetext}
	\begin{eqnarray}
		\label{hi}
		&&
		\frac{\partial
			\boldsymbol\Gamma}
		{\partial\cos\theta\,\partial\phi}
		=
		\frac{|\vec k|}{64\pi^2m_{\Lambda_b}^2}
		\Big[
		\boldsymbol N
		\left(
		1+p_z\cos\theta\,\vec s\cdot\hat r
		\right) 
		+
		p_z\boldsymbol\alpha\cos\theta
		+
		\boldsymbol\alpha\,\vec s\cdot\hat r
		+
		p_z\boldsymbol\beta\sin\theta\,\vec s\cdot\hat\phi
		-
		p_z\boldsymbol\gamma\sin\theta\,\vec s\cdot\hat\theta
		\Big] ,\label{eq:matga}
	\end{eqnarray}
	\end{widetext}
	where \(p_z\), chosen along the \(\hat z\)-axis, is the initial polarization
	of \(\Lambda_b\), \(\hat r\equiv\hat k\) with \(\hat k\) the outgoing
	direction of \(\Lambda\), and \(\cos\theta=\hat z\cdot\hat k\). The spin of
	\(\Lambda\) is described by \(\vec s=\chi^\dagger\vec\sigma_s\chi\), where
	\(\vec\sigma_s\) are the Pauli matrices in the spin space. The unnormalized
	flavor-space coupling matrices are directly related to the Lagrangian
	couplings by
	\(\boldsymbol N=\boldsymbol S\boldsymbol S^\dagger+
	\boldsymbol P\boldsymbol P^\dagger\),
	\(\boldsymbol\alpha=\boldsymbol P\boldsymbol S^\dagger+
	\boldsymbol S\boldsymbol P^\dagger\),
	\(\boldsymbol\beta=i\boldsymbol S\boldsymbol P^\dagger-
	i\boldsymbol P\boldsymbol S^\dagger\), and
	\(\boldsymbol\gamma=\boldsymbol S\boldsymbol S^\dagger-
	\boldsymbol P\boldsymbol P^\dagger\).
	The partial decay width for a specific neutral-\(D\) mode is obtained by
	projecting onto the corresponding flavor state,
	\(d\Gamma_D=\Tr(d\boldsymbol\Gamma\,\hat O_D)\), where \(\hat O_D\) was
	defined in Eq.~\eqref{flavor}.
	
	The normalized density matrix spans
	\({\cal H}_F\otimes{\cal H}_s\), with \({\cal H}_F\) the flavor and
	\({\cal H}_s\) the spin Hilbert space, and can be written generally as
	\begin{equation}
	\begin{aligned}
		\rho_{\vec k}
		=
		\frac14
		\bigg[
		&\mathbb 1_F\otimes\mathbb 1_s
		+
		\mathbb 1_F\otimes
		\left(
		\vec B_s\cdot\vec\sigma_s
		\right)
		\\
		&+
		\left(
		\boldsymbol B_D\cdot\boldsymbol\sigma
		\right)
		\otimes\mathbb 1_s
		+
		\sum_{i,j}
		C_{ij}\,
		\boldsymbol\sigma_i\otimes\sigma_s^j
		\bigg].\label{eq:matga1}
	\end{aligned}
	\end{equation}
	Here, \(\vec B_s\) and \(\boldsymbol B_D\) are the spin and flavor
	polarizations, respectively, and \(C_{ij}\) is the spin--flavor
	correlation matrix. \(i=x,y,z\) labels the flavor direction, and
	\(j=r,\theta,\phi\) that of the spin. On the one hand,
	\(\rho_{\vec k}\) can be obtained by projecting onto the momentum subspace
	of \(\rho_{J_z}\) with
    \begin{eqnarray}
    \rho_{\vec k} &=& 
    \frac{( 1+ p_z )}2 \hat O_{\vec k} \rho_{\frac{1}{2}} 
    \hat O_{\vec k}  
	  + \frac{( 1-  p_z )}2  \hat O_{\vec k} \rho_{-\frac{1}{2}} 
	  \hat O_{\vec k}   \,,
    \end{eqnarray}
    using the momentum projector
    \(\hat O_{\vec k}\), for which we refer the reader for more details to Appendix~\ref{appa}. On
    the other hand, \(\rho_{\vec k}\) is also directly related to the
    generalized partial-width matrix \(\boldsymbol\Gamma\). For instance, the
    spin polarization \(\vec B_s\) of \(\Lambda\) is given by tracing out the
    flavor subspace:
	\begin{equation}
	\begin{aligned}
		\vec B_s
		=
		\frac1n
		\Big[
		&\left(
		\Tr\left(\boldsymbol\alpha\right)
		+
		p_z\Tr\left(\boldsymbol N\right)\cos\theta
		\right)\hat r
		\\
		&-
		p_z\Tr\left(\boldsymbol\gamma\right)
		\sin\theta\,\hat\theta
		+
		p_z\Tr\left(\boldsymbol\beta\right)
		\sin\theta\,\hat\phi
		\Big],
	\end{aligned}
	\end{equation}
	where \(n\equiv\Tr(\boldsymbol N)+
	p_z\Tr(\boldsymbol\alpha)\cos\theta\). For completeness, we also show the
	results for the flavor polarization \(\boldsymbol B_D\) and the correlation
	matrix \(C_{ij}\):
	\begin{align}
		\boldsymbol B_D^i
		&=
		\frac{
		\Tr\left[
		\left(\boldsymbol N+p_z\cos\theta\,\boldsymbol\alpha\right)
		\boldsymbol\sigma_i
		\right]
		}{n},\nonumber 
		\\
		C_{ij}
		&=
		\frac1n
		\Tr\bigg\{
		\boldsymbol\sigma_i
		\bigl[
				\left(\boldsymbol N p_z\cos\theta+\boldsymbol\alpha\right)\delta_j^r \label{eq:matga2}
		\\
		&\qquad
		\qquad\quad+
		p_z\boldsymbol\beta\sin\theta\,\delta_j^\phi
		-
		p_z\boldsymbol\gamma\sin\theta\,\delta_j^\theta
		\bigr]
		\bigg\}.\nonumber 
	\end{align}
	with \(\delta_j^r\), \(\delta_j^\theta\), and \(\delta_j^\phi\) the
	Kronecker deltas in the spin-index space. Again, we stress that
	experimentally the flavor-\(y\) component cannot be tagged, and elements of
	the polarization vectors and the correlation matrix are directly related to
	the Lee--Yang parameters and the asymmetries
through the identities
	\begin{equation}
	\begin{aligned}
\Tr (
\boldsymbol \sigma _x  \boldsymbol \xi 
)
		&= 
			\Gamma_{D_1}\xi_{D_1}
			-
			\Gamma_{D_2}\xi_{D_2} \,,
	 \\
\Tr (
\boldsymbol \sigma _z  \boldsymbol \xi 
)
&=\Gamma_{\overline{D}^0}\xi_{\overline{D}^0}
	-
	\Gamma_{D^0}\xi_{D^0} \,.
	\end{aligned}
	\end{equation}
	These quantities can be determined numerically once the \(S\)- and \(P\)-wave
	amplitudes are specified.

    The size of spin--flavor entanglement in this momentum eigenstate is then
    quantified by~\cite{Wootters:1997id}
    \begin{align}
    {\cal C}_{\vec k} = \max(0,2\lambda_{\max}-\Tr{\cal R}),
    \end{align}
    with
    \({\cal R}=\sqrt{\sqrt{\rho_{\vec k}}(\boldsymbol\sigma_y\otimes\sigma_s^\theta)
    \rho_{\vec k}^\ast(\boldsymbol\sigma_y\otimes\sigma_s^\theta)\sqrt{\rho_{\vec k}}}\) and
    \(\lambda_{\max}\) the largest eigenvalue of \({\cal R}\).
    Here
    \({\cal C}_{\vec k}\) denotes the concurrence of the spin--flavor density
    matrix in the fixed-\(\vec k\) momentum-eigenstate subspace.    
    A direct Bell test is forbidden at colliders because the relevant spin and
    flavor components are not directly measurable~\cite{Abel:1992kz}.  Nevertheless, CHSH
    parameters can still diagnose entanglement and the collider-accessible
    construction is given by Eq.~\eqref{eq:chsh_phys} in Appendix~\ref{appa}.

\begin{figure}[h]
    \centering
    \includegraphics[width=0.8\columnwidth]{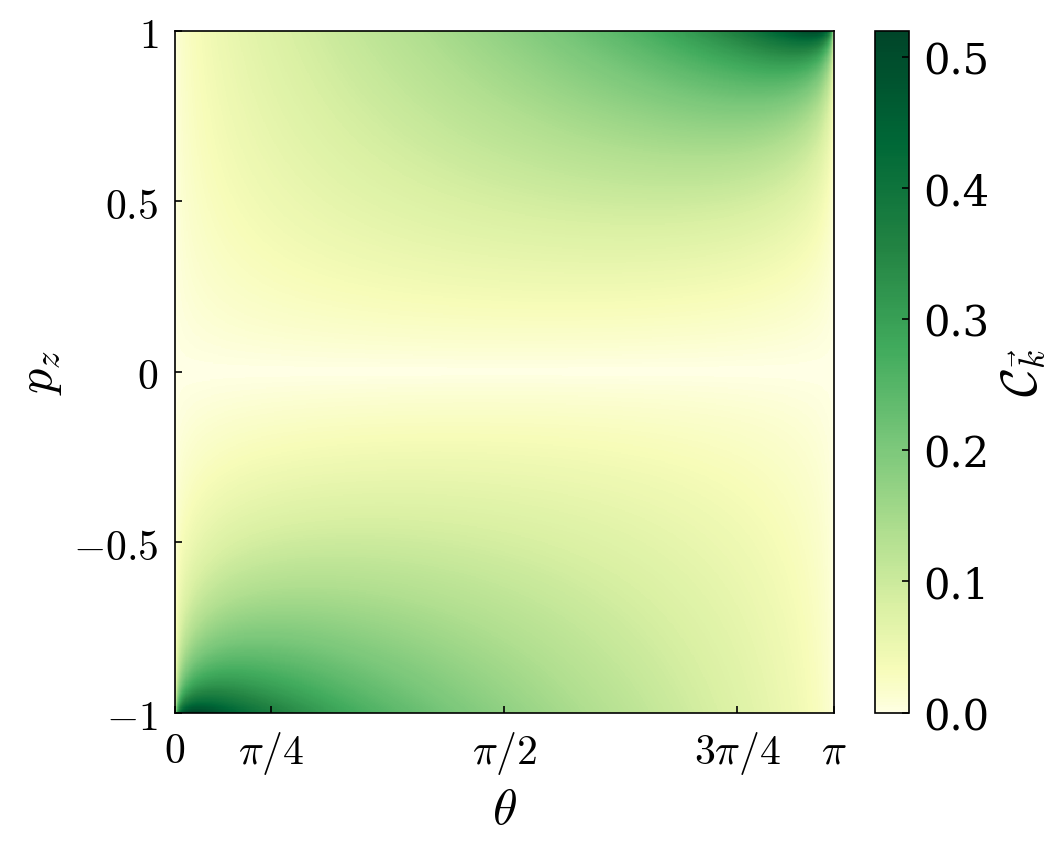}
    \caption{Theoretical prediction of the concurrence \({\cal C}_{\vec k}\) from the
    final-state spin--flavor entanglement in
    \(\Lambda_b\to\Lambda D\).}
    \label{fig:plotC}
    \end{figure}

    We show the theoretical prediction of \({\cal C}_{\vec k}\) in
    Figure~\ref{fig:plotC}, using recent theoretical \(S\)- and \(P\)-wave
    amplitude results from pQCD~\cite{Rui:2026ihu}. The plot shows that
    \({\cal C}_{\vec k}\) is rather sensitive to the initial polarization \(p_z\) of
    \(\Lambda_b\): (1) For \(p_z=0\), when the decaying \(\Lambda_b\) is unpolarized, it is clear from Eq.~\eqref{eq:matga}, or equivalently from Eqs.~\eqref{eq:matga1}--\eqref{eq:matga2}, that the state becomes separable such that \({\cal C}_{\vec k}=0\). As a consequence, the extraction of \(\gamma\) becomes inaccessible, which echoes   that $\beta_D$ and $\gamma_D$ cannot be measured at $p_z=0$. (2) For \(p_z\ne0\), non-vanishing spin--flavor entanglement can generically be induced, and maximal \({\cal C}_{\vec k}\) of \({\cal O}(0.5)\) can be achieved in a wide range of the decay angle \(\theta\), except when $\theta$ is close to 0 or $\pi$, i.e., the transverse production plane of $\Lambda_b$.
    
Historically, $p_z$ has been measured by various experiments at LEP and,
more recently, by the CMS and LHCb collaborations. These results are
summarized in Table~\ref{tab:pzexp}, and are all consistent with vanishing
$\Lambda_b$ polarization. The dominant uncertainty in these measurements is
statistical, and is expected to be substantially reduced, for example, by the
Tera-$Z$ runs of future lepton colliders such as
CEPC~\cite{CEPCStudyGroup:2018ghi} and FCC-ee~\cite{FCC:2018evy}. Such an
improved measurement of $p_z$ could in turn increase the accessible value of
${\cal C}_{\vec k}$ and thereby improve the precision of the $\gamma$
determination.

\vspace{0.2em}
\begin{table}[H]
    	\centering
    	\begin{tabular}{l||c|c}
    		\hline\hline
    		Exp. & $p_z$ & Ref.\\
    		\hline
    		ALEPH & $-0.23\pm0.23$  & {\text{\cite{ALEPH:1995aqx}}} \\
    		OPAL & $-0.56\pm0.19$   & {\text{\cite{OPAL:1998wmk}}} \\
    		DELPHI & $-0.49\pm0.35$ & {\text{\cite{DELPHI:1999hkl}}} \\
    		CMS & $0.00\pm0.08$     & {\text{\cite{CMS:2018wjk}}} \\
    		\hline
    		\multirow{3}{*}{LHCb} & $[-0.06, 0.05]_{\sqrt s=7\rm\,TeV}$ &  {\text{\cite{LHCb:2020iux}}} \\
    		& $[-0.04, 0.05]_{\sqrt s=8\rm\,TeV}$ & {\text{\cite{LHCb:2020iux}}} \\
    		& $[-0.01, 0.07]_{\sqrt s=13\rm\,TeV}$ & {\text{\cite{LHCb:2020iux}}} \\
    		\hline\hline
    	\end{tabular}
    	\caption{A summary of existing $p_z$ measurements at the lepton and the hadron colliders. The statistical and systematical uncertainties are summed in quadrature except in the LHCb case, the latter of which stands for the interval at the 68\% confidence level.}
    	\label{tab:pzexp}
    \end{table}
    
\section{Conclusion}

We have identified a new spin--flavor entanglement structure in
$\Lambda_b\to\Lambda D$ that correlates the $\Lambda$ spin and the neutral-$D$
flavor. The corresponding information is encoded in the decay rates and Lee--Yang parameters
of the $D=D^0,\overline D^0,D_1,D_2$ modes. This structure is both a
quantum feature of the final state and a probe of weak-phase
sensitivity: the weak-phase uncertainty scales as
$\sigma_\gamma\propto1/{\cal C}$, where ${\cal C}$ is the spin--flavor
concurrence. Thus, when the spin and flavor degrees of freedom become
separable, the extraction of $\gamma$ becomes ill-conditioned. In particular,
$\gamma$ cannot be extracted from branching fractions alone.

We also projected the same state onto local momentum wave packets and
constructed the corresponding spin--flavor density matrix. This local
description shows how the initial polarization and decay angle control the
observable concurrence. A realistic experimental implementation should use a
global likelihood for the rates, Lee--Yang parameters, efficiencies, and
covariances, but the central conclusion is basis independent: $\gamma$ is best
resolved when the final-state spin and flavor degrees of freedom are
strongly entangled.

\section*{Acknowledgements}

This work is supported in part by the National Natural Science Foundation of
China under Grant Nos. 12090064, 12375088, W2441004, 12547104, and
12575096, the Fundamental Research Funds for the Central Universities, and the
CAS One Hundred Talent Program.

	\clearpage 
	\onecolumngrid
	\appendix
	\section{
Eigenstate   decomposition 
	} \label{appa}

Before turning to the state decomposition, we recall the quark-level
Hamiltonian used in the main text:
\begin{equation}\label{app:heff_quark}
{\cal H}_{\rm eff}=\frac{G_F}{\sqrt2}
V_{cb}V_{us}^* 
\left[
C_1 Q_1^{(c)}+C_2 Q_2^{(c)} 
+\left|\frac{V_{ub}V_{cs}^*}{V_{cb}V_{us}^*}\right|e^{-i\gamma}
\left(C_1 Q_1^{(u)}+C_2 Q_2^{(u)}\right)\right]+{\rm h.c.}\, .
\end{equation}
With color indices \(\alpha,\beta\), the four-quark operators are
\begin{equation}\label{app:operators_buras}
\begin{aligned}
Q_1^{(c)}
&=
\left(\bar c_\alpha b_\beta\right)_{V-A}
\left(\bar s_\beta u_\alpha\right)_{V-A},
&
Q_2^{(c)}
&=
\left(\bar c_\alpha b_\alpha\right)_{V-A}
\left(\bar s_\beta u_\beta\right)_{V-A},
\\
Q_1^{(u)}
&=
\left(\bar u_\alpha b_\beta\right)_{V-A}
\left(\bar s_\beta c_\alpha\right)_{V-A},
&
Q_2^{(u)}
&=
\left(\bar u_\alpha b_\alpha\right)_{V-A}
\left(\bar s_\beta c_\beta\right)_{V-A}.
\end{aligned}
\end{equation}
Here
\((\bar q_\alpha q_\beta)_{V-A}\equiv
\bar q_\alpha\gamma^\mu(1-\gamma_5)q_\beta\). The \(S\)- and \(P\)-wave
amplitudes are defined through the momentum-space matrix element, with
\(\vec k\) the outgoing \(\Lambda\) momentum in the \(\Lambda_b\) rest frame
and \(\kappa=\sqrt{(E-m_\Lambda)/(E+m_\Lambda)}\):
\begin{equation}\label{app:sp_definition}
		{\cal M} _ {D}    ( J_z) = 	\langle \vec k,\vec s;\Lambda D|
	{\cal H}_{\rm eff}
	|\Lambda_b,J_z\rangle
	=
	i\,\overline u(\vec k,\vec s)
	\left(S_D-\frac{P_D}{\kappa}\gamma_5\right)
	u_b(J_z) .
\end{equation}
We start with the derivation of momentum eigenstates and then proceed to
angular-momentum eigenstates.

\subsection{Momentum eigenstates}
For a \(\Lambda_b\) state with fixed \(J_z\), the final state is projected onto momentum eigenstates as
	\begin{equation} \label{appe}
		| \psi_{J_z} ,\vec s , \vec k  \rangle =  
\hat O _{\vec k, \vec s }  {\cal H}_{eff} | \Lambda_b , J_z  \rangle \,,  
	\end{equation}
with the projector
	\begin{equation}
	\hat O _{\vec k ,\vec s  } =   
	\sum _{
		D
	} 
	| \vec k , \vec s ; \Lambda D \rangle
	\langle \vec k , \vec s ; \Lambda D | \,.
\end{equation} 
Here \(\vec s\) denotes the spin of \(\Lambda\), and the state
\(|\psi_{J_z},\vec s,\vec k\rangle\) has not been normalized.
It should be emphasized that
\(|\psi_{J_z},\vec s,\vec k\rangle\) is not itself an eigenstate of
\(J_z\). Rather, it is the fixed-\(\vec k\), fixed-\(\vec s\) component
projected out from the decay product of a fixed-\(J_z\) state. 
The density matrix 
for an initially polarized \(\Lambda_b\) is the incoherent mixture
 	\begin{equation}
	\begin{aligned}
		\rho ( \vec s  )
		= {}&
		\frac{1+p_z}{2}
		| \psi_{ \frac{1}{2}} , \vec s  ,\vec k  \rangle
		\langle
		\psi_{ \frac{1}{2}} , \vec s
		,\vec k  |
+ 
		\frac{1 - p_z}{2}
		|
		\psi_{  - \frac{1}{2}}   , \vec s
		,\vec k   \rangle
		\langle
		\psi_{  - \frac{1}{2}}
		, \vec s   ,\vec k   | \,.    
	\end{aligned}
	\end{equation}
The corresponding matrix element is
	\begin{equation}
		| \psi _{J_z} , \vec s , \vec k \rangle \langle
		\psi _{J_z} , \vec s , \vec k | =
		\sum _{
			D , D'
		}
		| \vec k , \vec s ; \Lambda D \rangle
		\langle \vec k , \vec s ; \Lambda D' |
		{\cal M} _ {D }    ( J_z)
		{\cal M} _ {D' }    ( J_z)  ^* \,.   
		\end{equation}
		We note that
		\begin{equation}
		{\cal M} _ {D }  (J_z  )
		{\cal M} _ {D' }  ^* (J_z  )    =
		\,\overline{u} ( \vec s \, )
		\left(
		S _ { D}  -
		\frac{
			P _ {D }
		}{\kappa}
		\gamma_5
		\right)
		u_b (J_z)
		\overline{u} _b( J_z )
		\left(
		S _ {D'}^*  +
		\frac{
			P _ {D' } ^*
		}{\kappa}
		\gamma_5
		\right)
		u(\vec s\, )
		\end{equation}
		has the same structure as in the usual derivation~\cite{LeeYang:1957}. For example,
		\begin{equation}
		{\cal M} _ {D }
		{\cal M} _ {D' }  ^*
		=
		\frac14\,
		\mathrm{Tr}\Bigg[
		(\slashed p_\Lambda + m_\Lambda)(1+\gamma_5 \slashed s)
		\left(
		S _ { D}  -
		\frac{
			P _ {D }
		}{\kappa}
		\gamma_5
		\right)
		(\slashed p_b+m_{\Lambda_b})(1+\gamma_5 \slashed s_b)
		\left(
		S _ {D'}^*  +
		\frac{
			P _ {D' } ^*
		}{\kappa}
		\gamma_5
		\right)
		\Bigg] \,,
	\end{equation}
where $s _b = ( 0, 0 ,0,J_z)$. 
	The trace can be evaluated straightforwardly using \textsc{Mathematica}.
	
	To express the result in matrix form, we adopt the basis
	\begin{equation}
		| \vec k , \vec s ; \Lambda D^0 \rangle
		= ( 1 ,0  ) ^T ,~~~
		| \vec k , \vec s ; \Lambda \overline{D}^0 \rangle
		= ( 0 ,1 ) ^T \,.
	\end{equation}
	Then the flavor density matrix elements are
\begin{equation}
	\rho
	=
	\begin{pmatrix}
		{\cal M}_{D^0}{\cal M}_{D^0}^*
		&
		{\cal M}_{D^0}{\cal M}_{\overline{D}^0}^*
		\\
		{\cal M}_{\overline{D}^0}{\cal M}_{D^0}^*
		&
		{\cal M}_{\overline{D}^0}{\cal M}_{\overline{D}^0}^*
	\end{pmatrix}.
\end{equation} 
	In terms of Eq.~\eqref{M_kin},
$ 
	\rho \propto
	\boldsymbol{{\cal M} }
	\boldsymbol{{\cal M} }^\dagger
	$ and, since $d\Gamma \propto \rho$, Eq.~\eqref{hi} follows immediately.

    From the density matrix setup above, the Clauser--Horne--Shimony--Holt (CHSH)
	parameter ${\cal B}$~\cite{CHSH:1969,Horodecki:1995} can be computed from the Horodecki bound,
	\({\cal B}\leq2\sqrt{\mu_1^2+\mu_2^2}\), where \(\mu_i^2\) are the eigenvalues
	of \({\mathbf C}{\mathbf C}^{\,T}\), ordered as
	\(\mu_1^2\geq\mu_2^2\geq\mu_3^2\).  Since the flavor-\(y\) direction is experimentally inaccessible, the physical CHSH parameter is
	obtained from the \(x\) and \(z\) flavor components:
\begin{equation}
	\begin{aligned}
			\max\left[{\cal B}_{\rm phys}\right]
		&=
		2\sqrt{
			\sum_{i=x,z}
			\left(
			\hat{\mathbf C}\hat{\mathbf C}^{\,T}
			\right)_{ii}
		}
		 \\
		&= 
		\frac{2}{1+p_z\alpha_{\rm av}\cos\theta}
		\sqrt{
			\left(
			p_z\cos\theta\,R_{12}
			+
			\alpha_{12}
			\right)^2
			+
			\left(
			p_z\cos\theta\,R_F
			+
			\alpha_F
			\right)^2
			+
			p_z^2\sin^2\theta
			\left(
			\beta_{12}^2+\beta_F^2+\gamma_{12}^2+\gamma_F^2
			\right)
		},\label{eq:chsh_phys}
	\end{aligned}
\end{equation}
	with 
	$ 
	\alpha_{\rm av}
	=
	\frac12
	\left[
	\left(1+R_F\right)\alpha_{\overline{D}^0}
	+
	\left(1-R_F\right)\alpha_{D^0}
	\right]
	$. 
	At \(p_z=0\), \({\mathbf C}{\mathbf C}^{\,T}\) becomes a rank-one
	matrix, such that \(\mu_2^2=0\) and \({\cal B}\leq2\). In the same limit, the
	concurrence also vanishes.  The maximal value obtained by varying the
	Lee--Yang parameters is
	\begin{equation}
		\max_{\xi}
		\left[
		\mu_1^2+\mu_2^2
		\right]
		=
		1
		+
		\frac{
			p_z^2\sin^2\theta
		}{
			1-p_z^2\cos^2\theta
		}.
	\end{equation}
A simple example that saturates this bound for \(p_z=1\) at
\(\theta=\pi/2\) is given by
\(S_{D^0}=iP_{\overline{D}^0}\) and
\(P_{D^0}=S_{\overline{D}^0}=0\). 
	
	\subsection{
Angular momentum eigenstates 
	}

Let $\vec k=|\vec k|(\sin\theta\cos\phi,\sin\theta\sin\phi,\cos\theta)$,
where $(\theta,\phi)$ are the spherical coordinates of the $\Lambda$ momentum.
The helicity angular-momentum eigenstates are defined by
\begin{equation}\label{helicity_state_app}
	|J_z,\lambda;\Lambda D\rangle
	=
	\frac{1}{2\pi}
	\int d\Omega\,
	|\vec k,\lambda;\Lambda D\rangle\,
	e^{iJ_z\phi}
	d^{1/2}_{J_z\lambda}(\theta) .
\end{equation}
Here $\lambda$ is the helicity of $\Lambda$. The helicity basis is related to
the spin basis by
\begin{equation}\label{kinetic}
	|\vec k,\lambda=\pm 1/2;\Lambda D\rangle
	=
	|\vec k,\vec s=\pm \hat k;\Lambda D\rangle ,
\end{equation}
That is, in a helicity eigenstate, the spin vector $\vec s$ is aligned or
anti-aligned with the momentum direction $\hat k$.

The fixed-$J_z$ helicity states form a complete basis in the $\Lambda D$
subspace. We therefore introduce
\begin{equation}\label{OJz_def}
	\hat O_{J_z}
	=
	\sum_{\lambda=\pm 1/2}
	\sum_{D}
	\frac{1}{{\cal N}_2}
	|J_z,\lambda;\Lambda D\rangle
	\langle J_z,\lambda;\Lambda D| ,
\end{equation}
where
$
{\cal N}_2
\equiv
\langle J_z,\lambda;\Lambda D
|J_z,\lambda;\Lambda D\rangle
$
is the normalization factor. In the fixed-$J_z$ $\Lambda D$ subspace,
$\hat O_{J_z}$ acts as the identity:
\begin{equation}\label{He_cal}
	\langle \vec k,\vec s;\Lambda D|
	{\cal H}_{\rm eff}
	|\Lambda_b,J_z\rangle
	=
	\langle \vec k,\vec s;\Lambda D|
	\hat O_{J_z}{\cal H}_{\rm eff}
	|\Lambda_b,J_z\rangle .
\end{equation}
The unnormalized final state is then
\begin{align}
	|\psi, {J_z}\rangle_{\rm un}
	&=
	\hat O_{J_z}{\cal H}_{\rm eff}|\Lambda_b,J_z\rangle
	 =
	\sum_{\lambda=\uparrow,\downarrow}
	\sum_{D}
	c_{\lambda D}
	|J_z,\lambda;\Lambda D\rangle .
	\label{eq:psi_un_sum}
\end{align}
Here $\uparrow$ and $\downarrow$ denote $\lambda=+1/2$ and $-1/2$,
respectively. For fixed $J_z$, all information about the decay
$\Lambda_b\to\Lambda D$ is encoded in the four spin--flavor coefficients
\begin{equation}
	c_{\lambda D}
	\equiv
	\langle J_z=\lambda,\lambda;\Lambda D|
	{\cal H}_{\rm eff}
	|\Lambda_b,J_z=\lambda\rangle . 
	\label{eq:c_lambdaD_def}
\end{equation} 
The coefficients $c_{\lambda D}$ can be obtained from
Eq.~\eqref{He_cal} by choosing $\vec k=|\vec k|\hat z$ and
$\vec s=\lambda\hat k$. Explicitly,
\begin{align}
	\langle \vec k,\vec s;\Lambda D|
	{\cal H}_{\rm eff}
	|\Lambda_b,J_z\rangle
	&=
	\frac{1}{{\cal N}_2}
	\langle \vec k,\vec s;\Lambda D
	|J_z,\lambda;\Lambda D\rangle
	\langle J_z,\lambda;\Lambda D|
	{\cal H}_{\rm eff}
	|\Lambda_b,J_z\rangle
	\nonumber\\
	&=
	\delta_{J_z\lambda}\,
	c_{\lambda D} .
	\label{eq:c_from_momentum}
\end{align}
In deriving the last line, we used
$
\langle \vec k,\vec s;\Lambda D
|J_z,\lambda;\Lambda D\rangle
=
{\cal N}_2 d_{J_z\lambda}^{1/2}(0)
$
and
$d^{1/2}_{J_z\lambda}(0)=\delta_{J_z\lambda}$.  
Equations~\eqref{eq:c_from_momentum} and
\eqref{app:sp_definition} determine the coefficients
$c_{\lambda D}$ in terms of the partial-wave amplitudes $S_D$ and $P_D$,
reproducing Eq.~\eqref{heli_exp}. 
 
With $\alpha_D$ and $\Gamma_D$, upper and lower bounds on ${\cal C}$ 
can be obtained from  
\begin{equation}\label{interval}
	{\cal C}_{\min,\max}
	=
	\frac{2}{\Gamma_{\overline{D}^0}+\Gamma_{D^0}}
	\left[
	\Gamma_{\overline{D}^0}\Gamma_{D^0}
	-\frac{\left(\Gamma_{D_1}-\Gamma_{D_2}\right)^2}{4}
	-\frac{\left(I_{+}\pm I_{-}\right)^2}{4}
	\right]^{1/2},
\end{equation}
where the upper and lower signs correspond to ${\cal C}_{\min}$ and
${\cal C}_{\max}$, respectively, and
\begin{equation}
	I_{\pm}\equiv
	\left[
	\Gamma_{\overline{D}^0}\Gamma_{D^0}
	(1\pm\alpha_{\overline{D}^0})(1\pm\alpha_{D^0})
	-
	\frac{
		\left[
		\Gamma_{D_1}(1\pm\alpha_{D_1})
		-
		\Gamma_{D_2}(1\pm\alpha_{D_2})
		\right]^2
	}{4}
	\right]^{1/2}.
\end{equation}

\section{Binary Lee--Yang operators in the sampling}
\label{app:lee_yang_binary}
 
The binary nature of the Lee--Yang operators follows from the spin algebra.  We use
the commutator relations~\cite{spin}
\begin{equation}
	[ s_i, s_j ] = i \epsilon_{ijk} s_k \,,~~~
	[k_i, J_j ] = i \epsilon_{ijk} k _k \,,~~~
	[s_i, J_j ] = i \epsilon_{ijk} s_k\,. 
\end{equation}
The local spin components acting on the momentum-helicity state
\(| \vec{k}, \lambda ; \Lambda D \rangle\) satisfy
\begin{equation}
	2s_\theta 
	| \vec{k} , \pm \frac{1}{2} \rangle
	=
	| \vec{k} , \mp  \frac{1}{2} \rangle
	\,,~~~
	2s_\phi 
	| \vec{k} , \pm \frac{1}{2} \rangle
	= 
	\pm 
	i
	| \vec{k} , \mp  \frac{1}{2} \rangle \,,~~~
	2s_r 
	| \vec{k} , \pm \frac{1}{2} \rangle
	= 
	\pm 
	| \vec{k} ,    \pm \frac{1}{2} \rangle\,,
\end{equation} 
where the label \(\Lambda D\) has been suppressed.  The usual
spin-\(1/2\) angular-momentum operators act on the
\(|J_z,\lambda;\Lambda D\rangle\) basis as
\begin{equation}
	J_x
	| J_z ,  \lambda \rangle 
	= \frac{1}{2}
	|  -J_z ,  \lambda \rangle 
	\,,~~~
	J_y
	|   J_z ,  \lambda \rangle 
	= 
	i J_z 
	|  -J_z ,  \lambda \rangle  \,,~~~
	J_z 
 | J_z,  \lambda \rangle 
	= 
J_z
 |   J_z,  \lambda \rangle  \,,
\end{equation}
again suppressing the \(\Lambda D\) label.  Combining these relations with
the Wigner rotation between momentum-helicity states
in Eq.~\eqref{helicity_state_app}   gives
\begin{equation}
\begin{aligned}
	\hat\alpha |J_z,\lambda\rangle
 =
	2\lambda |J_z,\lambda\rangle,
~~~ 
	\hat\gamma |J_z,\lambda\rangle
	 =
	|J_z,-\lambda\rangle,
	~~~
	\hat\beta |J_z,\lambda\rangle
	 =
	-2i\lambda |J_z,-\lambda\rangle .
\end{aligned}
\end{equation}
The Lee–Yang operators do not change $J_z$ or $J$, since they are rotational scalars satisfying $[\hat{\xi},\vec J]=0$. 
Therefore, in the \(\Lambda\)-helicity basis,
\(\hat\alpha\), \(\hat\gamma\), and \(\hat\beta\) are represented by
\(\sigma_z\), \(\sigma_x\), and \(-\sigma_y\), respectively.  In particular,
\begin{equation}
	\hat\xi^2=1,\qquad
[\hat\alpha,\hat\beta]=2i\hat\gamma. 
\end{equation}
Each channel Lee--Yang operator   has eigenvalues \(\pm1\).  
 
Given an observable $\hat O$, the statistical uncertainty is governed by
$
\sigma_O =
\sqrt{
	\left(
	\langle \hat O^2\rangle
	-
	\langle \hat O\rangle^2
	\right)/N
},
$
where $N$ is the number of samples. Suppose that there are $N_D$ events for
$\Lambda_b\to \Lambda D$ for a specific flavor $D$. If the Lee--Yang operators
are measured directly with equal statistics, the uncertainties are
\begin{equation}\label{b6}
	\sigma_{\xi_D}^{\rm dir}
	=
	\sqrt{
		\frac{3(1-\xi_D^2)}{N_D}
	}\, .
\end{equation}
In reality, the spin measurements are inferred from the decay
$\Lambda\to p\pi^-$, with
$
\alpha_\Lambda \langle \vec s\rangle = \langle \hat p\rangle ,
$
where $\alpha_\Lambda$ is the up-down asymmetry of $\Lambda\to p\pi^-$,
and $\hat p$ is the proton momentum direction in the $\Lambda$ rest frame.
Projecting onto the $D$ channel in Eq.~\eqref{hi}, with the normalization
$\langle \hat O_D\rangle=1$, we obtain
\begin{equation}
	\alpha_D
	=
	\left\langle \hat O_D\,\frac{3p_r}{\alpha_\Lambda}\right\rangle,
	\qquad
	\gamma_D
	=
	\left\langle \hat O_D\,\frac{-12p_\theta}{\pi p_z\alpha_\Lambda}\right\rangle,
	\qquad
	\beta_D
	=
	\left\langle \hat O_D\,\frac{12p_\phi}{\pi p_z\alpha_\Lambda}\right\rangle .
\end{equation}
The corresponding uncertainties are
\begin{equation} \label{uncertainties}
	\begin{aligned}
		\sigma_{\alpha_D}
		&=
		\frac{1}{|\alpha_\Lambda|}
		\sqrt{
			\frac{3-\alpha_\Lambda^2\alpha_D^2}{N_D}
		},
		~~~
		\sigma_{\gamma_D}
		=
		\frac{4}{\pi |p_z\alpha_\Lambda|}
		\sqrt{
			\frac{
				3
				-\left(\frac{\pi}{4}p_z\alpha_\Lambda\gamma_D\right)^2
			}{N_D}
		},
		~~~
		\sigma_{\beta_D}
		=
		\frac{4}{\pi |p_z\alpha_\Lambda|}
		\sqrt{
			\frac{
				3
				-\left(\frac{\pi}{4}p_z\alpha_\Lambda\beta_D\right)^2
			}{N_D}
		}.
	\end{aligned}
\end{equation} 
The $1/|p_z\alpha_\Lambda|$ scaling of the transverse uncertainties reflects
that $\beta_D$ and $\gamma_D$ enter Eq.~\eqref{hi} only through the
$p_z\sin\theta$ terms and are inferred through the spin-analyzing power
$\alpha_\Lambda$ of $\Lambda\to p\pi^-$.

\section{Sensitivity to \(\gamma\) near the small-concurrence limit}
\label{app:gamma_density_matrix}

This appendix summarizes how the weak-phase sensitivity is controlled by the
spin--flavor concurrence.  Flavor-tagged rates and Lee--Yang parameters fix
the norms and directions of the helicity spinors
\(\chi_D,\chi_{\overline{D}}\), but not their relative phase.  The
\(D_1,D_2\) modes and their CP conjugates supply the interference needed to
determine that phase together with \(\gamma\).

In general, the physics is invariant under a $U(2)$ rotation of the helicity
spinors, $\chi_{\overline D,D}\to U\chi_{\overline D,D}$. Since
$\chi_{\overline D,D}$ are determined only up to a relative phase, we can choose
the helicity basis such that
\begin{equation}
	\chi_{\overline{D}}
	=
	\begin{pmatrix}
		u'\\
		0
	\end{pmatrix},
	\qquad
	\chi_D
	=
	e^{i\phi_u}
	\begin{pmatrix}
		u\\
		v
	\end{pmatrix},
	\qquad
	u',u,v>0,
\end{equation}
where $\phi_u$ is the strong phase.  Here \(u\) and \(v\) are the positive real
components of \(\chi_D\) parallel and perpendicular to
\(\chi_{\overline{D}}\), respectively.  We work with normalized amplitudes,
\({\cal N}=1\).  In this basis,
\begin{equation}
	u'u
	=
	\left|
	 \chi _{\overline D} ^\dagger  \chi _{ {D}}
	\right|
	=
	\frac{1}{2}\sqrt{1-R_F^2-{\cal C}^2}, 
	\qquad
	u'v
	= \left| \chi _{\overline D} ^T i \sigma _y \chi _{ {D}} \right| 
=
	\frac{{\cal C}}{2}.
\end{equation}
The second relation shows that the small-concurrence limit corresponds to a
small perpendicular component \(v\).

We decompose the \(\Lambda_b\) density matrix in flavor space as
\begin{equation}
\begin{aligned}
	\rho_{J_z}
	&= 
	\begin{pmatrix}
		\chi_{\overline D}\chi_{\overline D}^{\dagger}
		&
		e^{-i\gamma}\chi_{\overline D}\chi_D^{\dagger}
		\\
		e^{i\gamma}\chi_D\chi_{\overline D}^{\dagger}
		&
		\chi_D\chi_D^{\dagger}
	\end{pmatrix} = 
	\begin{pmatrix}
		\rho_0+\rho_z
		&
		\rho_x-i\rho_y
		\\
		\rho_x+i\rho_y
		&
		\rho_0-\rho_z
	\end{pmatrix}
\end{aligned}
\end{equation}
where each \(\rho_i\) is a \(2\times2\) matrix in helicity space.  The
CP-conjugate density matrix is decomposed analogously, with blocks
\(\bar\rho_i\).  Since \(\rho_y\) corresponds to the unphysical flavor
direction \(D^0\pm i\overline{D}^0\), it is not directly observable
experimentally.  We therefore compare \(\rho_x\) with the CP-conjugate block
after applying the helicity flip,
$
	\widetilde\rho_x
	\equiv
	\sigma_x\bar\rho_x\sigma_x .
$ 

In the basis above, the off-diagonal helicity entries are
\begin{equation}
	(\rho_x)_{12}
	=
	\frac{e^{-i(\phi_u + \gamma)}}{2}\,u'v ,
	\qquad
	(\widetilde\rho_x)_{12}
	=
	\frac{
e^{-i(\phi_u - \gamma) 	} 
}{2}\,u'v .
\end{equation}
The ratio of the two entries removes the common strong phase \(\phi_u\) and gives
\begin{equation}
	\gamma 
	=
	-\frac{1}{2}
	\arg
	\left[
		\frac{(\rho_x)_{12}}{(\widetilde\rho_x)_{12}}
	\right]
	\quad {\rm mod}\ \pi .
\end{equation}
This modulo-\(\pi\) ambiguity is embedded in the parametrization through the
simultaneous shift \(\gamma\to\gamma+\pi\) and
\(\phi_u\to\phi_u+\pi\), analogous to the usual \(B\to DK\) ambiguity.
The magnitude of each off-diagonal coherence is fixed by the concurrence, given by 
\begin{equation}
	|(\rho_x)_{12}|
	=
	|(\widetilde\rho_x)_{12}|
	=
	\frac12 u'v
	=
	\frac{{\cal C}}{4}.
\end{equation}
Let \(V_{\rm off}\) denote the covariance matrix of
\(\{\mathrm{Re}(\rho_x)_{12},\mathrm{Im}(\rho_x)_{12},
\mathrm{Re}(\widetilde\rho_x)_{12},\mathrm{Im}(\widetilde\rho_x)_{12}\}\).
Gaussian error propagation for the phase ratio gives
\begin{equation}\label{eq:gamma_off_uncertainty}
	\sigma_{\gamma,{\rm off}}^2
	=
	\frac14\,
	\mathbf g^T V_{\rm off}\,\mathbf g
	\equiv
	\frac{\sigma_\xi^2}{{\cal C}^2},
	\qquad
	\mathbf g
	=
	\begin{pmatrix}
	-16\,\mathrm{Im}(\rho_x)_{12}/{\cal C}^2\\
	16\,\mathrm{Re}(\rho_x)_{12}/{\cal C}^2\\
	16\,\mathrm{Im}(\widetilde\rho_x)_{12}/{\cal C}^2\\
	-16\,\mathrm{Re}(\widetilde\rho_x)_{12}/{\cal C}^2
	\end{pmatrix}.
\end{equation}
Here \(\sigma_\xi\) is a finite effective Lee--Yang uncertainty after
covariance projection.  The factors \(16/{\cal C}^2\) in \(\mathbf g\) come
from differentiating the phase of a complex number with
\(|(\rho_x)_{12}|=|(\widetilde\rho_x)_{12}|={\cal C}/4\), so that
\(1/|(\rho_x)_{12}|^2=16/{\cal C}^2\); the prefactor \(1/4\) is from the
square of the \(1/2\) in the phase-ratio extraction.

The diagonal entries also contain the weak phase, but only through cosine
constraints.  In the same basis,
\begin{equation}
	(\rho_x)_{11}
	=
	u'u\cos(\gamma+\phi_u),
	\qquad
	(\widetilde\rho_x)_{11}
	=
	u'u\cos(\phi_u - \gamma).
\end{equation}
Since \(u'u\) is fixed by \(R_F\) and \({\cal C}\), the diagonal entries give
the branch-dependent solution
\begin{equation} 
	\gamma_{\rm diag}
	=
	\frac{1}{2}
	\left[
	\eta_1 \arccos\left(\frac{(\rho_x)_{11}}{u'u}\right)
	+
	\eta_2 \arccos\left(\frac{(\widetilde\rho_x)_{11}}{u'u}\right)
	\right]
	\ {\rm mod}\ \pi,
	\quad
	\eta_1,\eta_2=\pm1 .
\end{equation}
Without knowing the branch signs $\eta_1$ and $\eta_2$, the diagonal
constraints give several discrete solutions for $\gamma$ in
$0\leq\gamma<\pi$.  Averaging over these unresolved branches would give
$\pi/2$, independently of the true value of $\gamma$, and therefore has no
physical meaning as an extraction of the weak phase.  Thus the diagonal
information alone cannot determine $\gamma$ unambiguously.  The effective precision therefore depends on the off-diagonal
phase information, and we parametrize the branch-conditioned contribution by
\begin{equation}
	\sigma_{\gamma,{\rm diag}}^{\rm eff}
	=
	\frac{f_{\rm br}(R_F,{\cal C})}{{\cal C}}\,
	\sigma'_\xi .
\end{equation}
Here \(\sigma'_\xi\) is the corresponding effective Lee--Yang uncertainty, and
\(f_{\rm br}\) encodes the branch geometry, local slopes, and experimental
covariance, which is in general nonzero.  The inverse-variance combination with
Eq.~\eqref{eq:gamma_off_uncertainty} gives
\begin{equation}
	\sigma_\gamma^{\rm full}
	\simeq
	\frac{1}{{\cal C}}
	\left[
		\frac{1}{\sigma_\xi^2}
		+
		\frac{1}{
			f_{\rm br}^2(R_F,{\cal C})(\sigma'_\xi)^2
		}
	\right]^{-1/2}.
\end{equation}
Thus
$ 
	\sigma_\gamma^{\rm full}
	\propto
 {\cal C}^{-1},$
at $ 
	{\cal C}\to0 .
$ 
This scaling should not be extrapolated beyond the Gaussian regime, where the
likelihood becomes a mixture over the possible diagonal branches.


\begin{thebibliography}{99}

\bibitem{Afik:2020onf}
Y.~Afik and J.~R.~M.~de Nova,
Eur. Phys. J. Plus \textbf{136}, 907 (2021),
arXiv:2003.02280.

\bibitem{ATLAS:2023fsd}
G.~Aad \textit{et al.} (ATLAS Collaboration),
Nature \textbf{633}, 542 (2024),
arXiv:2311.07288.

\bibitem{CMS:2024pts}
A.~Hayrapetyan \textit{et al.} (CMS Collaboration),
Rept. Prog. Phys. \textbf{87}, 117801 (2024),
arXiv:2406.03976.

\bibitem{Gu:2025rg}
J.~Gu, S.~J.~Lin, D.~Y.~Shao, L.~T.~Wang, and S.~X.~Yang,
arXiv:2510.13951.

\bibitem{Han:2024kct} 
T.~Han, M.~Low, N.~McGinnis and S.~Su,
JHEP \textbf{05}, 081 (2025) 
arXiv:2412.21158.
 


\bibitem{Fang:2026ddi}
Y.~J.~Fang, A.~Bhoonah, K.~Cheng, T.~Han, Y.~Liu and H.~Zhang,
arXiv:2604.11887.

\bibitem{Cheng:2025xux} 
K.~Cheng and B.~Yan,
Phys. Rev. Lett. \textbf{135},   1 (2025)
arXiv:2501.03321.

\bibitem{Cao:2025qua}
Q.~H.~Cao, G.~Li, X.~K.~Wen and B.~Yan,
arXiv:2509.18276.


\bibitem{BESIII:2018cnd}
M.~Ablikim \textit{et al.} (BESIII Collaboration),
Nature Phys. \textbf{15}, 631 (2019),
arXiv:1808.08917.

\bibitem{Perotti:2018wxm}
E.~Perotti, G.~F\"aldt, A.~Kupsc, S.~Leupold, and J.~J.~Song,
Phys. Rev. D \textbf{99}, 056008 (2019),
arXiv:1809.04038.

\bibitem{Wu:2024bne}
S.~Wu, C.~Qian, Q.~Wang and X.~R.~Zhou,
Phys. Rev. D \textbf{110}, 054012 (2024)
arXiv:2406.16298.


\bibitem{Du:2024sly}
Y.~Du, X.~G.~He, C.~W.~Liu and J.~P.~Ma,
Eur. Phys. J. C \textbf{85}, 1255 (2025),
arXiv:2409.15418.

\bibitem{Lin:2025eci}
S.~J.~Lin, M.~J.~Liu, D.~Y.~Shao and S.~Y.~Wei,
JHEP \textbf{11}, 082 (2025),
arXiv:2507.15387.

\bibitem{Chen:2026oaf}
C.~Chen and J.~J.~Xie,
arXiv:2603.24011.

\bibitem{Feng:2025ryr}
H.~L.~Feng, H.~Tang, W.~Z.~Guo and Q.~Qin,
Phys. Rev. D \textbf{112}, 036020 (2025),
arXiv:2504.15798.

\bibitem{STAR:2025njp}
B.~E.~Aboona \textit{et al.} (STAR Collaboration),
Nature \textbf{650}, 65 (2026),
arXiv:2506.05499.

\bibitem{Oliva:2026qsc}
L.~Oliva, Q.~Wang, and X.~N.~Wang,
arXiv:2603.10427.

\bibitem{Belle:2007EPR}
A.~Go \textit{et al.} (Belle Collaboration),
Phys. Rev. Lett. \textbf{99}, 131802 (2007),
arXiv:quant-ph/0702267.

\bibitem{LeeYang:1957}
T.~D.~Lee and C.~N.~Yang,
Phys. Rev. \textbf{108}, 1645 (1957).

\bibitem{Cabibbo:1963yz}
N.~Cabibbo,
Phys. Rev. Lett. \textbf{10}, 531 (1963).
 
\bibitem{Kobayashi:1973fv}
M.~Kobayashi and T.~Maskawa,
Prog. Theor. Phys. \textbf{49}, 652 (1973).

\bibitem{PDG:2024}
S.~Navas \textit{et al.} (Particle Data Group),
Phys. Rev. D \textbf{110}, 030001 (2024).

\bibitem{Brod:2013sga}
J.~Brod and J.~Zupan,
JHEP \textbf{01}, 051 (2014),
arXiv:1308.5663.

\bibitem{Gronau:1991dp}
M.~Gronau and D.~Wyler,
Phys. Lett. B \textbf{265}, 172 (1991).

\bibitem{Gronau:1991dk}
M.~Gronau and D.~London,
Phys. Lett. B \textbf{253}, 483 (1991).

\bibitem{Atwood:1996ci}
D.~Atwood, I.~Dunietz, and A.~Soni,
Phys. Rev. Lett. \textbf{78}, 3257 (1997),
arXiv:hep-ph/9612433.

\bibitem{Atwood:2000ck}
D.~Atwood, I.~Dunietz, and A.~Soni,
Phys. Rev. D \textbf{63}, 036005 (2001),
arXiv:hep-ph/0008090.

\bibitem{Giri:2003ty}
A.~Giri, Y.~Grossman, A.~Soffer, and J.~Zupan,
Phys. Rev. D \textbf{68}, 054018 (2003),
arXiv:hep-ph/0303187.

\bibitem{LHCb:2021sqa}
R.~Aaij \textit{et al.} (LHCb Collaboration),
JHEP \textbf{12}, 141 (2021),
arXiv:2110.02350.

\bibitem{BelleBelleII:2024}
I.~Adachi \textit{et al.} (Belle and Belle II Collaborations),
JHEP \textbf{10}, 143 (2024),
arXiv:2404.12817.

\bibitem{Buchalla:1995vs}
G.~Buchalla, A.~J.~Buras and M.~E.~Lautenbacher,
Rev. Mod. Phys. \textbf{68}, 1125 (1996),
arXiv:hep-ph/9512380.
 

\bibitem{Geng:2022osc}
C.~Q.~Geng, X.~N.~Jin, C.~W.~Liu, Z.~Y.~Wei and J.~Zhang,
Phys. Lett. B \textbf{834}, 137429 (2022) arXiv:2206.00348.

\bibitem{spin}
A. McKerrell, {\it  Nuovo Cim} {\bf 34}, 1289 (1964).

\bibitem{Wootters:1997id}
W.~K.~Wootters,
Phys. Rev. Lett. \textbf{80}, 2245 (1998).
 
\bibitem{LHCb:2016qpe}
R.~Aaij \textit{et al.} (LHCb Collaboration),
Phys. Rev. Lett. \textbf{118}, 052002 (2017);
Erratum: Phys. Rev. Lett. \textbf{119}, 169901 (2017),
arXiv:1612.05140.

\bibitem{Vecchi:2018bjg}
S.~Vecchi (LHCb Collaboration),
EPJ Web Conf. \textbf{192}, 00024 (2018).

\bibitem{LHCb:2019fns}
R.~Aaij \textit{et al.} (LHCb Collaboration),
Phys. Rev. D \textbf{100}, 031102 (2019),
arXiv:1902.06794.

\bibitem{Rui:2026ihu}
Z.~Rui, Z.~T.~Zou, Y.~Li and Y.~Li,
arXiv:2604.17877.

\bibitem{Hocker:2001xe}
A.~Hocker, H.~Lacker, S.~Laplace and F.~Le Diberder,
Eur. Phys. J. C \textbf{21}, 225-259 (2001)
arXiv:hep-ph/0104062;
J.~Charles \textit{et al.} [CKMfitter Group],
Eur. Phys. J. C \textbf{41},   1-131 (2005) 
arXiv:hep-ph/0406184.

\bibitem{ALEPH:1995aqx}
D.~Buskulic \textit{et al.} [ALEPH],
Phys. Lett. B \textbf{365}, 437-447 (1996).

\bibitem{OPAL:1998wmk}
G.~Abbiendi \textit{et al.} [OPAL],
Phys. Lett. B \textbf{444}, 539-554 (1998)
arXiv:hep-ex/9808006.

\bibitem{DELPHI:1999hkl}
P.~Abreu \textit{et al.} [DELPHI],
Phys. Lett. B \textbf{474}, 205-222 (2000).

\bibitem{CMS:2018wjk}
A.~M.~Sirunyan \textit{et al.} [CMS],
Phys. Rev. D \textbf{97}, no.7, 072010 (2018)
[arXiv:1802.04867 [hep-ex]].

\bibitem{LHCb:2020iux}
R.~Aaij \textit{et al.} [LHCb],
JHEP \textbf{06}, 110 (2020)
arXiv:2004.10563.

\bibitem{CEPCStudyGroup:2018ghi}
J.~B.~Guimar{\~a}es da Costa \textit{et al.} [CEPC Study Group],
arXiv:1811.10545.

\bibitem{FCC:2018evy}
A.~Abada \textit{et al.} [FCC],
Eur. Phys. J. ST \textbf{228},  261-623 (2019).
 

\bibitem{Abel:1992kz}
S.~A.~Abel, M.~Dittmar, and H.~K.~Dreiner,
Phys. Lett. B \textbf{280}, 304 (1992).

\bibitem{CHSH:1969}
J.~F.~Clauser, M.~A.~Horne, A.~Shimony, and R.~A.~Holt,
Phys. Rev. Lett. \textbf{23}, 880 (1969).

\bibitem{Horodecki:1995}
R.~Horodecki, P.~Horodecki, and M.~Horodecki,
Phys. Lett. A \textbf{200}, 340 (1995).
		
		
		
	\end{thebibliography}
\end{document}